# Room-temperature memristive switching between charge density wave states

R. Venturini[1,2,*], M. Rupnik[3,4], J. Gašperlin[1,3], J. Lipič[1,3], P. Šutar[1], Y. Vaskivskyi[1,3], F. Ščepanović[1,3,4], D. Grabnar[1], D. Golež[1,3,*], D. Mihailovic[1,3,4]

[1]Jozef Stefan Institute, 1000 Ljubljana, Slovenia

[2]PSI Center for Photon Science, Paul Scherrer Institute, 5232 Villigen PSI, Switzerland

[3]Faculty of Mathematics and Physics, University of Ljubljana, 1000 Ljubljana, Slovenia

[4]CENN Nanocenter, 1000 Ljubljana, Slovenia

* Correspondence to: rok.venturini@psi.ch, denis.golez@ijs.si

**Control over charge density wave (CDW) states is technologically promising for the development of ultra-efficient memory devices. However, using electrical pulses for non-volatile resistance switching involving CDW states has so far been limited to cryogenic temperatures. In this work, we investigate a recently discovered layered semiconductor $EuTe_4$, which exhibits the coexistence of distinct CDW orders. We report that electrical pulses can be used for excitation to non-equilibrium, yet stable electronic states across a broad temperature range from 6 K to 400 K. We find that switching occurs through a non-thermal pathway and is reversible via a thermal erase procedure. The resistance of the new electronic state is tunable by the pulse voltage, so the device acts as a memristor. Calculations show fixed bilayer CDW, whereas CDW in single layers shows bistability due to weak Eu-Te links. Low-voltage, fast, and energy-efficient CDW switching holds potential for memristor applications.**

## Introduction

The rapid growth of artificial neural networks has drastically increased computational demands, driving the need for neuromorphic hardware that could implement fast and energy-efficient artificial neurons and synapses[1]. The memristor is a key component in bio-inspired computing architectures as it enables the storage and manipulation of information via non-volatile control of channel resistance[2]. Traditional memristors rely on slow and often unreliable transport of ions through solids[3,4]. Other approaches, such as phase-change memory[5–7] and Mott switches[8–10] rely on transforming the atomic lattice via Joule heating across a structural phase transition. This results in energy-inefficient and abrupt switching, deviating from the ideal linear resistance programming[6,11]. A recent approach involves resistance switching between distinct charge density wave (CDW) states, as demonstrated in $1T$-$TaS_2$[12–15]. As the switching is driven by electron reconfiguration, such mechanism was shown to be ultrafast[16], ultraefficient[12,17,18], and has high endurance[17]. However, the non-volatile operation is limited to cryogenic temperatures[19].

The study of novel CDW states is also of fundamental interest, as optical and electrical pulse excitation provide a promising route for accessing hidden charge-ordered states that form in out-of-equilibrium conditions[20–28]. The lifetime of such states is generally limited to the picosecond timescale[21,26,27,29], as it is exceedingly rare to obtain a sufficiently high energy barrier between distinct CDW states for long-lived metastability. Electrical pulse switching in 1*T*-$TaS_2$ has remained an exception to the rule, where stability on laboratory timescales and at cryogenic temperatures is provided via the topologically protected network of CDW domains and domain walls[30]. Retaining novel charge-ordered states through metastability is not only valuable for potential applications, but it also offers a rare opportunity to study thermally inaccessible quantum states.

Here, we investigate $EuTe_4$, a recently discovered layered semiconductor with intricate CDW physics[31–37]. Its electronic states close to the Fermi surface originate from both single and bilayer square tellurium sheets (Figs. 1a and 1b) that are susceptible to Fermi surface nesting[31,32,38,39]. The incommensurate unidirectional CDW along the *b* axis (Fig. 1d) is present at room temperature[32] and persists to temperatures as high as 652 K[33,40]. Recent works have also uncovered additional in-plane CDW domains oriented along the *a* axis[33,34] and domains with tilted CDW with respect to the atomic lattice[34]. Further complexity arises from the out-of-plane degree of freedom as the bilayer sheets are aligned, while the monolayer is staggered with respect to the bilayers (Fig. 1a). As incommensurate CDWs in Te monolayers and bilayers have slightly different periods, the interaction between the two creates a unique large moiré CDW superstructure[37]. While band calculations predict metallic behavior in the high-temperature state[31,34], the co-existence and interaction between the multiple CDW orders is reflected in the anomalous transport properties of $EuTe_4$, showing semiconducting behavior with a giant thermal hysteresis around room temperature. Competition and coexistence of distinct CDW orders make $EuTe_4$ a promising candidate for exploring long-lived CDW metastabilities.

Recently, ultrafast laser pulses have been used for non-volatile switching of the $EuTe_4$ resistance at room temperature[41]. While optical pulses of moderate fluence kept the resistance of the material within the hysteresis loop, high-fluence pulses could be used for switching outside the thermal hysteresis loop. The detection of a second harmonic signal has shown that optically induced change in material resistance is related to the CDW order, however, the exact nature of the novel order is still to be clarified. Direct electrical pulse manipulation of the CDW could open a new pathway to control CDW states at room temperature and enable the fabrication of a fast and ultraefficient memristor.

## Results

### Resistance switching at room temperature

In this work, we report on electrical pulse switching to novel CDW states in $EuTe_4$ that are exceptionally stable and therefore allow non-volatile memristive operation even at room temperature. Our devices (schematically shown in Fig. 1c) are fabricated from exfoliated $EuTe_4$ flakes using optical lithography (see Methods). Fig. 1d shows the resistance measurement of device D1, which exhibits a typical semiconducting and hysteretic behavior of $EuTe_4$ crystals. We observe that by applying a current pulse to the cooling branch at 300 K, the resistance of the device increases and is stable on the laboratory timescales. By increasing the pulse magnitude up to 2.7 V, a large change in device resistance can be induced (the switching direction is indicated with a white arrow in Fig. 1e). As shown in Fig. 1e, the change of resistance is erased by performing a thermal cycle up to 430 K. After electrical pulsing we perform a full thermal cycle (Figs. 1d and 1e) and observe that the resistance value of the electrically switched state lies well outside of the hysteresis loop. This means that the new state induced by an electrical pulse is a thermally inaccessible state, as it cannot be reached via thermal cycling. Next, we demonstrate the remarkable stability of the novel state by applying a 2.0 V pulse to the cooling branch at an even higher temperature of 350 K (inset to Fig. 1e). Despite the high temperature, the novel state is highly stable during the next 15 hours (the duration of the experiment).

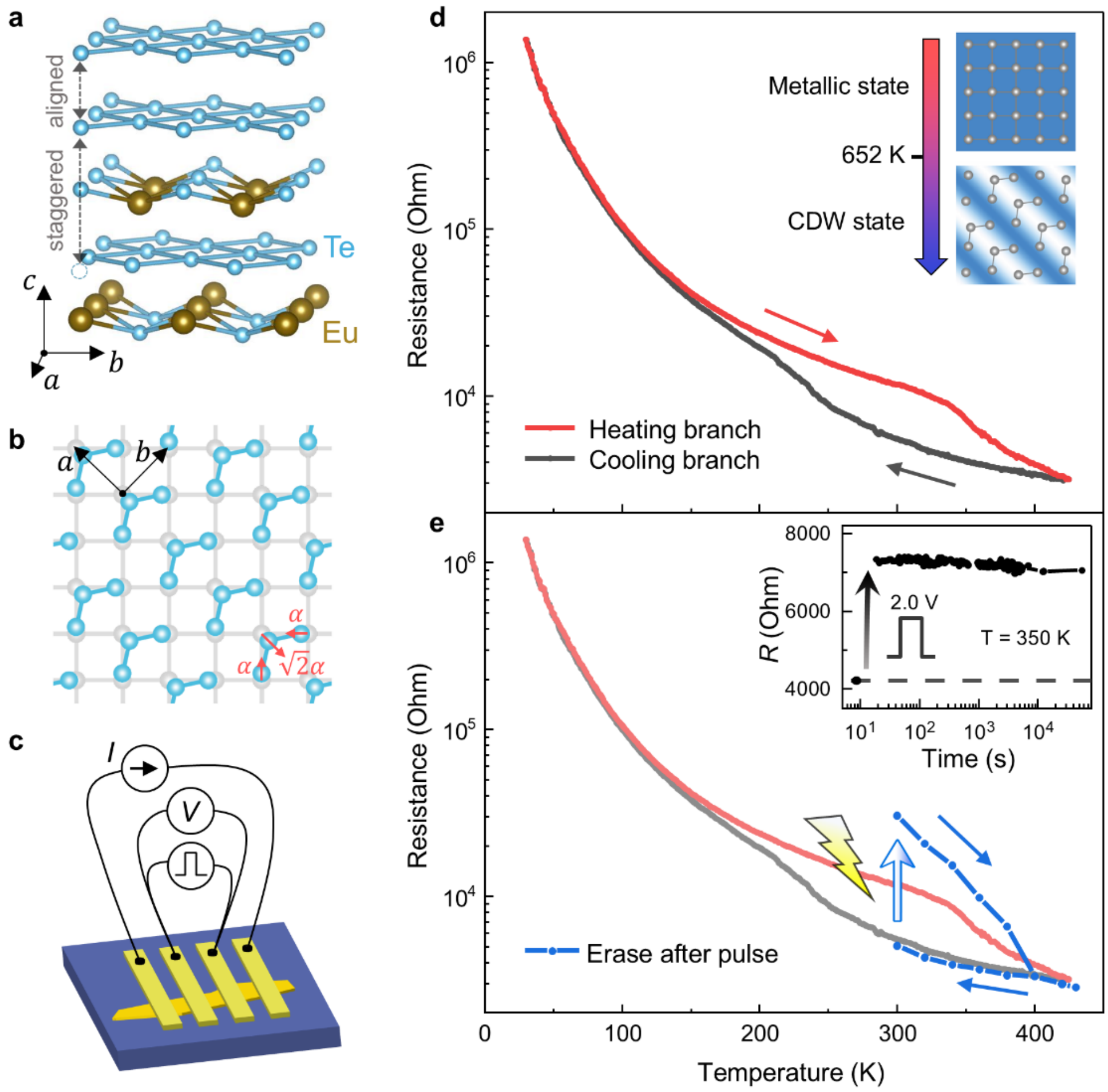

**Fig. 1: $EuTe_4$ and electrical pulse switching of device D1.** **a**, Crystal structure of $EuTe_4$. **b**, A square lattice of Te atoms (gray) and a distorted lattice hosting the CDW state (blue). $\alpha$ denotes a parameter of the lattice distortion inside the CDW state. **c**, Schematic of a $EuTe_4$ electrical device with a current source, voltmeter, and voltage pulse source connected to the device. **d**, Thermal hysteresis of the $EuTe_4$ device during the cooling (black) and heating (red) process. The inset shows an illustration of the material phase diagram with the charge density modulation indicated by white stripes. **e**, Thermal erase cycle (blue) following the resistance switching by a 2.7 V pulse at 300 K (the direction of switching indicated with a white arrow). The inset shows the stability of the switched state at 350 K after applying a 2.0 V pulse.

In Fig. 2 we present the details of the electrical switching behavior at 300 K. Fig. 2a shows a systematic study of switching by increasing the electrical pulse voltage starting at either the cooling or heating branch. We observe distinct behavior for low-voltage and high-voltage pulses, separating the response into two regimes. In the low-voltage regime, resistance increases when starting at the cooling branch and decreases when starting from the heating branch. In the high-voltage regime, the resistance appears independent of the starting state and continually increases with increasing voltage. This indicates that in the low-voltage regime, the CDW order retains a partial history of the starting state, which is fully erased when switching in a high-voltage regime. As for device D1, we observe that switching is possible to thermally inaccessible states that are outside of the thermal hysteresis loop. Interestingly, the resistance of the $EuTe_4$ device gradually changes with increasing voltage which contrasts with the abrupt switching of CDW states observed in 1*T*-$TaS_2$[12]. We confirm that the resistance change is a single shot effect determined by the pulse amplitude, by applying a sequence of pulses with the same magnitude (see Supplementary Fig. 1).

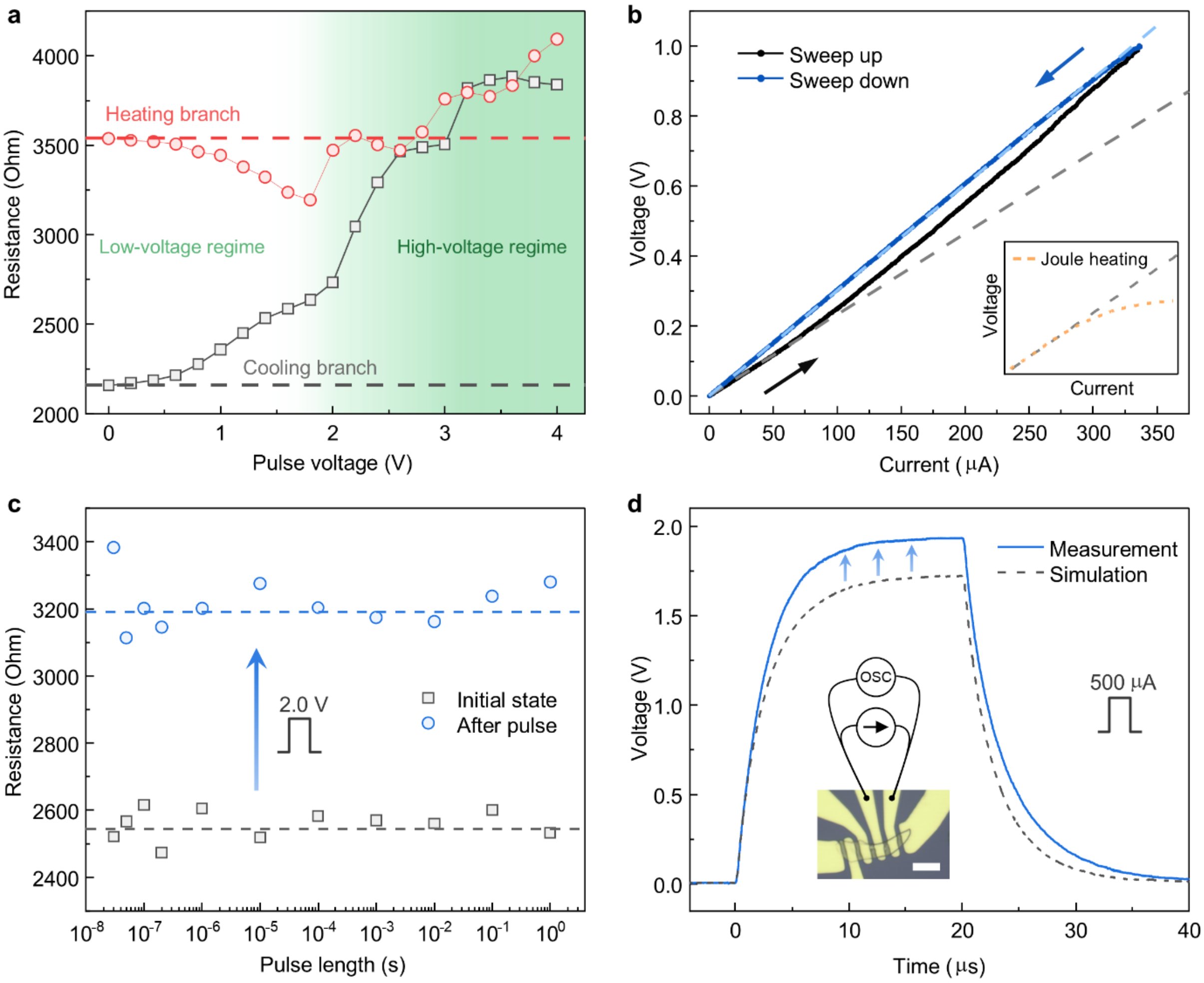

**Fig. 2: Electrical switching of device D2 at 300 K.** **a**, Device resistance after applying voltage pulses of increasing magnitude, starting from either the cooling branch (black) or the heating branch (red). A distinct response is observed for low-voltage and high-voltage regimes (green shading). **b-d**, Electrical pulse switching from the cooling branch. **b**, Voltage measured during the current pulses by incrementally increasing (black) and decreasing (blue) pulse magnitude. The dashed lines are a linear fit to the low-voltage region. The inset shows a schematic of a sub-linear $I$-$V$ response of a semiconductor with a significant change in temperature due to Joule heating. A distinct response of a semiconducting $EuTe_4$ device indicates minimal change in temperature during pulse application. **c**, Change in resistance after applying 2.0 V voltage pulses of varying lengths, demonstrating switching reproducibility and pulse duration independence. **d**, Direct temporal observation of the switching process during a 500 μA current pulse (blue) and simulated response that assumes no resistance change during the current pulse (black dashed line). The change in resistance is observed already within the initial microsecond rise time, which sets the upper limit on device switching speed. Insert: a microscope image of the device with connections for a current source and an oscilloscope (scalebar: 10 μm).

Fig. 2b presents a current-voltage ($I$-$V$) characteristic of the $EuTe_4$ device which is measured using pulsed current to minimize Joule heating. The device response is linear at low voltages and then starts to deviate above the linear extrapolation already at a low voltage of 0.15 V (see Supplementary Note 8), consistent with the onset of resistance switching shown in Fig. 2a. After reaching 1.0 V, the current magnitude is gradually decreased with voltage~~,~~ showing a linear response. This contrasts with Joule heating response of a semiconductor which would result in a sub-linear $I$-$V$ curve as schematically shown in the inset to Fig. 2b. The absence of any sub-linear behavior in the $I$-$V$ response indicates that switching is non-thermal and performed without significant heating of the $EuTe_4$ device.

To study the switching speed, we apply single voltage pulses of different lengths (30 ns to 1 s) and measure resistance before and after switching. After each switching, the state is thermally erased by heating to 425 K. As shown in Fig. 2c, we observe that switching is remarkably consistent across the entire range of pulse lengths and multiple switching and erasing cycles. The independence of switching on the pulse length demonstrates that switching is fast and done by the leading edge of the pulse, with the rise time which is presently limited to 30 ns by our circuit (see Supplementary Fig. 5).

Having shown that switching can be triggered with very short pulses, we next aim to explore how fast the system settles into the final state. To observe a change in resistance on short timescales, we use short current pulses and measure the voltage across the device during the pulse with an oscilloscope (Fig. 2d). We observe that the high resistance state is already reached during the current pulse's microsecond rise time as seen by comparing the observed voltage trace to the simulated response which assumes no resistance switching during the pulse (see Supplementary Note 7). Presently, the circuit rise time prevents us from exploring even shorter timescales and allows us to set the upper boundary for reaching the final state to the sub-microsecond timescale; however, we anticipate that the process is significantly faster.

**Temperature dependence of electrical switching**

Next, we explore the electrical switching behavior over a broad range of temperatures between 6 K and 300 K. We note that the hysteresis curve of D2 (Fig. 3a) exhibits some variation in shape compared to device D1. Such variability is also observed in the literature, where the exact shape of the hysteresis appears to differ between crystals used in separate works[32,38,40] and could be due to a small level of impurities of the Eu source[40], difference in device thickness, or local strain. As shown in Fig. 3a, we use low-voltage electrical pulses for which the change in resistance is moderate and at room temperature within the hysteresis loop. One might expect that the change of resistance after applying an electrical pulse would be smaller at lower temperatures, where the hysteresis shrinks. However, we find the opposite: at low temperatures, even low voltage pulses switch the sample to non-thermal electronic states outside of the thermal hysteresis. Note that at low temperatures, the voltage pulse drives significantly lower current due to a higher device resistance, further reducing any potential Joule heating effects. Furthermore, data presented in Fig. 3a demonstrate that switching is reversible throughout the explored temperature range. Fig. 3b shows that the switched states are remarkably stable throughout the explored temperature range.

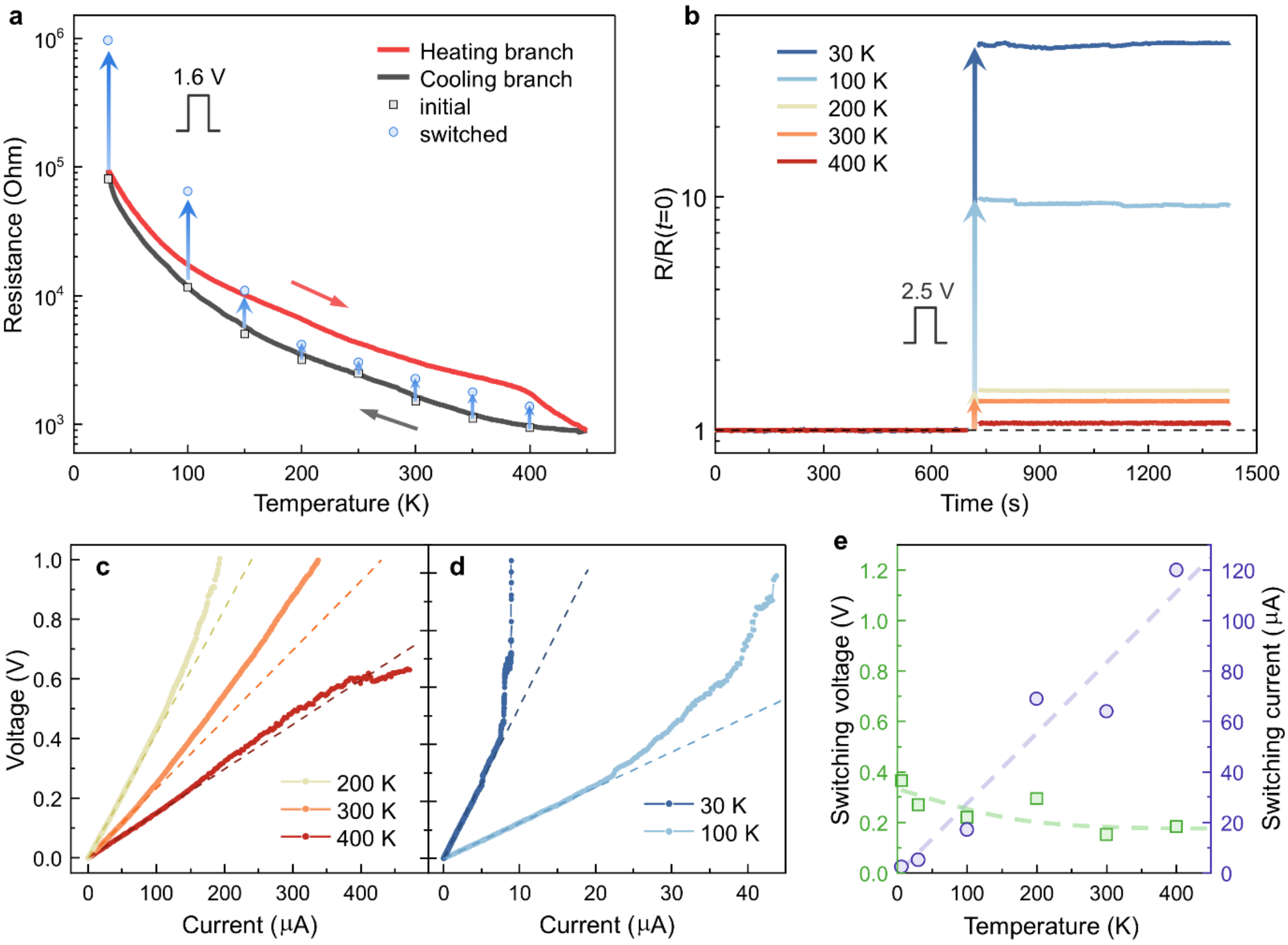

**Fig. 3: Temperature dependence of switching in device D2 from the cooling branch.** **a**, Electrical switching with a low magnitude 1.6 V voltage pulse at different temperatures. **b**, Time dependence of device resistance before and after switching with a 2.5 V pulse at different temperatures. **c**,**d**, Pulsed *I*-*V* characteristics for high- and low-temperature regimes, respectively. **e**, Temperature dependence of the threshold switching voltage (green) and current (purple).

Pulsed *I-V* measurements shown in Figs. 3c and 3d reveal a similar response at all explored temperatures: a linear response at low voltages is followed by an upward deviation which signifies resistance switching. The *I-V* characteristic obtained at 400 K deviates from this behavior at high currents, which we attribute to Joule heating. At low temperatures, the switching appears in larger discrete steps, signifying a larger resistance change with a small change in voltage. To determine whether switching in $EuTe_4$ flakes is driven by an electric field or charge injection, we track switching parameters for different temperatures, leveraging the significant change of the device resistance with temperature. To determine the switching voltage and current quantitatively, we fit the low-voltage part of the *I-V* curve and record the threshold values where the measured *I-V* curves deviate from the extrapolated response (see Supplementary Note 8). As shown in Fig. 3e, the threshold current for switching is significantly reduced as the temperature is lowered, while the threshold voltage remains at a similar value throughout the explored temperature range. A mechanism primarily driven by charge injection would be expected to occur at a constant threshold current, as has been observed in 1*T*-$TaS_2$[12]. In contrast, our observation of a constant threshold voltage strongly indicates that the electric field is the primary driver of the switching.

## Discussion

Electrical pulsing is highly efficient in the CDW state manipulation, as the observed switching voltage is surprisingly small, especially considering the large device size with a 2 μm gap between electrodes. For the 30 ns pulses applied at room temperature, we calculate that the total energy deposited in the device is only 13 pJ, which limits the temperature increase during the pulse to 6.6 K (see Supplementary Note 5). We summarize additional observations that support the non-thermal nature of switching. (1) Switching is independent of the pulse length, inconsistent with Joule-heating-driven switching[6,42]. (2) An *I-V* characteristic of a semiconductor affected by Joule heating has a sub-linear response due to decreasing resistance with increasing temperature[42]. We do not observe such a response in our measurements. (3) In a thermal scenario, the pulse-dissipated power should increase with lowering temperature to provide sufficient heat for switching. In contrast, we observe that the switching current is reduced by lowering the temperature. (4) A thermal heating cycle reduces the device resistance, regardless of the initial state (see Supplementary Fig. 1). Electrical switching has the opposite effect of increasing the device resistance.

The non-thermal, low-voltage, and gradual switching observed in $EuTe_4$ devices is markedly different from existing resistance switching devices[3,8,9], offering promising characteristics for novel memory technologies like memristors. Interestingly, the switching mechanism also differs significantly from the CDW switching in 1*T*-$TaS_2$, where charge injection induces an abrupt transformation of the long-range ordered CDW state into a textured phase with domain walls, accompanied by a drop in resistance[12,30,43]. Unlike 1*T*-$TaS_2$, where switching in a device appears along a narrow channel[44], the overall large increase in resistance in $EuTe_4$ devices indicates that the majority of the device is involved in the switching process (see Supplementary Note 5 for further details). Furthermore, the established mechanism of electric-field-induced CDW sliding, observed in one- and two-dimensional CDW materials[45–49] can't explain the switching behavior in $EuTe_4$, as it lacks the non-linear *I-V* response before the switching threshold. We can exclude mechanisms such as Zener tunnelling and avalanche breakdown as observed in dielectrics or Mott insulators, as these processes are volatile, accompanied by a sub-linear IV response, and result in a decrease in device resistance[9].

Interestingly, resistance switching is independent of the pulse polarity, and we could not perform reversible switching simply by reversing the electric field (see Supplementary Note 10). This observation rules out switching due to ion migration, as observed for memristive switching in oxides[9]. While $EuTe_4$ hosts a polar CDW order[41], the absence of a bipolar response is incompatible with a simple in-plane reversal of polar order, distinguishing it from the established switching of polarization in ferroelectric materials[50,51]. Calculating the switching electric field at room temperature from the device dimensions and switching voltage, we obtain the threshold values of $E_{D1} \sim 0.86$ kV/cm, and $E_{D2} \sim 0.75$ kV/cm (see Supplementary Notes 8 and 9 for details). These values are significantly lower than the ~1 MV/cm electric field values needed to induce Zener tunelling[9] or reverse the out-of-plane polarization in van der Waals material $WTe_2$[52]. The switching field is also lower than the avalanche breakdown field of 1-10 kV/cm observed in narrow-gap semiconductors[53]. Although the sliding of the CDW is a volatile process and it can't explain the switching observed in $EuTe_4$, the threshold field values for sliding observed in 1*T*-$TaS_2$ at room temperature are, notably, on a similar order of magnitude of ~1 kV/cm[49].

Our electrical switching results show several parallels with recent optical manipulation experiments in $EuTe_4$ done at room temperature[41]. This indicates that non-volatile CDW switching is intrinsically linked to the unique CDW physics of $EuTe_4$. On the cooling branch of the thermal hysteresis, both electrical pulses and optical excitation increase the resistance. On the heating branch, the response is similarly non-monotonic: lower intensity stimuli (weaker electric fields or lower laser fluences) decrease the resistance, while higher intensity stimuli cause it to increase again. Furthermore, much like high-fluence optical experiments, strong electrical pulses can induce a non-thermal state far outside the equilibrium hysteresis loop. In contrast to optical switching, only low voltage is required for device operation, allowing us to clearly isolate the non-thermal pathway and demonstrate energy-efficient memristive control.

Recent X-ray diffraction experiments have indicated that the interlayer interaction of the incommensurate CDW orders could be relevant, as it creates competing moiré domains with different out-of-plane stacking orderings already in equilibrium[37]. The manipulation of such domains aligns well with our switching observations, supporting a complex free-energy landscape of CDW states in $EuTe_4$, characterized by numerous local minima separated by energy barriers sufficiently high to enable non-volatile operation at room temperature.

To theoretically explore the microscopic origin of these free-energy local minima, we extend Peierls' theory of charge-density waves to several coupled layers as relevant for $EuTe_4$. We begin with a monolayer model and later include various interlayer stacking configurations to understand the energetics of excited states (see Methods). First, our calculation confirms that for the monolayer, consistent with experiments[31,36], the Lindhard response exhibits a peak very close to $\mathbf{q} = (1/3a^*, 1/3a^*)$, where $a^*$ is the reciprocal lattice constant of the monolayer unit cell (i.e., rotated by 45° w.r.t. the unit cell of $EuTe_4$ and scaled accordingly[54]), see Supplementary Fig. 12. For simplicity, we assume a commensurate wavevector and parameterize the distortion using the parameter $\alpha$ (see Fig. 1b), which is determined self-consistently (see Methods). In contrast to previous density-functional theory calculations[31], the band structure of the monolayer CDW state is now completely gapped (see Supplementary Fig. 13), and its energy landscape has minima at positive ($\alpha > 0$) and negative ($\alpha < 0$) distortions, which are symmetry-related and thus degenerate (see Fig. 4a).

Next, we examine the energetics of interlayer stacking. $EuTe_4$ exhibits two relevant stacking configurations: aligned bilayers and staggered monolayer Te sheets, see Fig. 1a. Since the bilayer and monolayer are separated by Eu-Te layers, their coupling acts as a weak link. In our calculations, we assume a fixed antiferroelectric bilayer structure (green and orange layers in Figs. 4b and 4c) and vary the monolayer distortion $\alpha$, which is our main order parameter. There is a stable configuration for $\alpha > 0$ and a metastable configuration for $\alpha < 0$ (see Figs. 4a–c). These configurations are similar to states proposed for the upper and lower hysteresis, respectively, within the Landau-Ginzburg description[32]. Moreover, our microscopic description enables an estimate of the potential barrier to several meV. This is consistent with experimental observations that weak excitations can switch the state, but the precise height of the barrier is challenging to estimate due to uncertainties in the microscopic parameters. A comparison of the band structures for the two states shows that the state with $\alpha > 0$ always exhibits a larger band gap than the state with $\alpha < 0$ (see Supplementary Fig. 13). Consequently, the resistivity of the former is also expected to be higher, which would be consistent with experimental observations. Beyond this transition, we found other metastable minima at higher energies associated with various interlayer distortions (not shown). Additional experimental input is needed to determine which of these configurations are experimentally relevant.

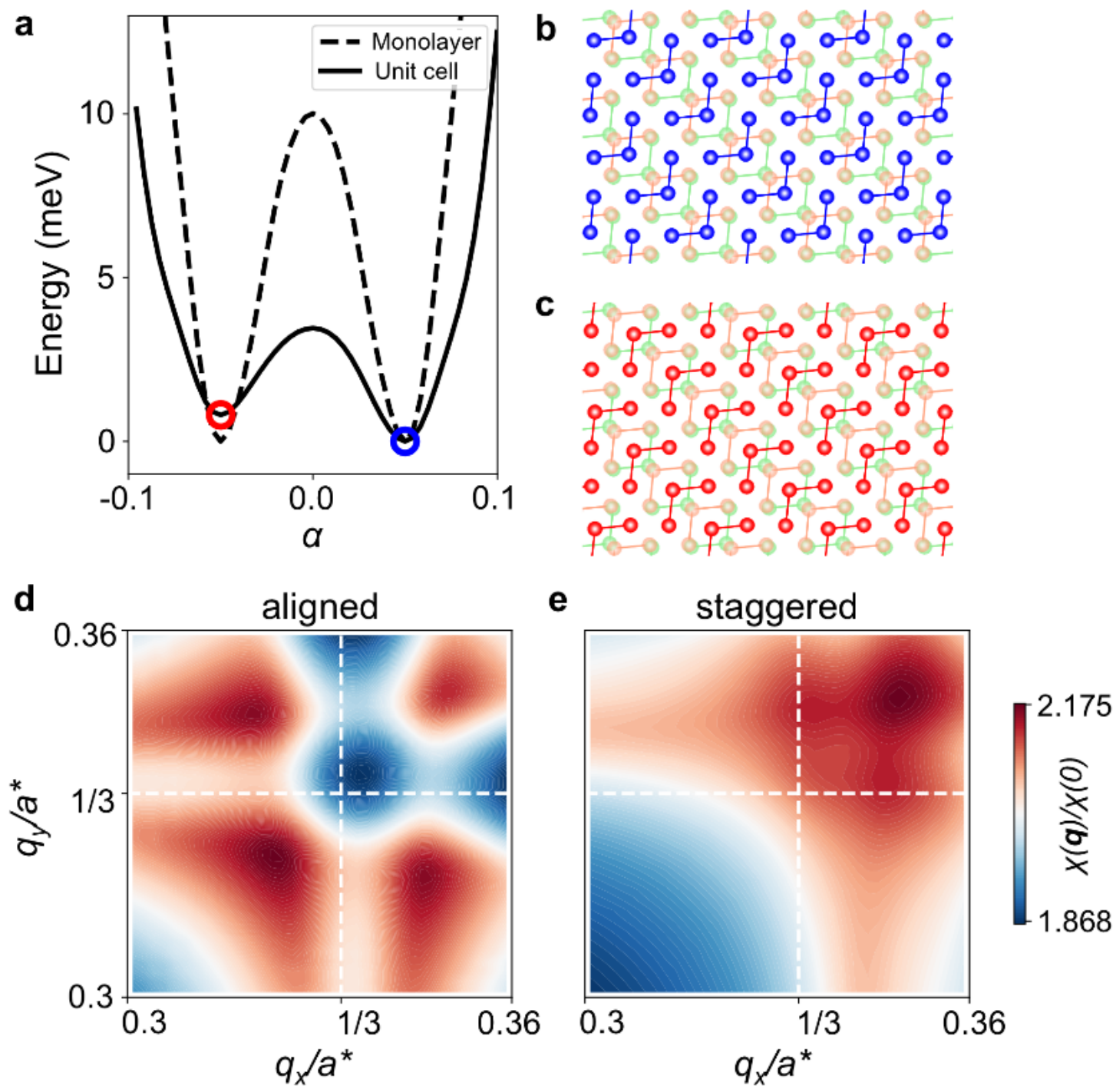

**Fig. 4: Microscopic modeling of intralayer and interlayer CDW patterns.** **a**, The energy landscape as a function of the lattice distortion $\alpha$ of a monolayer (dashed line) or the unit cell composed of three layers with a fixed bilayer (full line). The blue and red circles mark the ground and metastable states, respectively. **b**,**c**, Real-space atomic configurators where the top (bottom) layer of the bilayer is shown in transparent orange (green). **d**,**e**, The Lindhard response ($\chi$) for an undistorted aligned and staggered bilayer geometry.

Finally, we show that the interlayer stacking is coupled with the in-plane order by considering the Lindhard response of aligned or staggered bilayers' geometry, see Figs. 4d and 4e. In both cases, the original peak close to $\mathbf{q} = (1/3a^*, 1/3a^*)$ is split, showing maxima for collinear and noncolinear wave vectors[36]. This means that the coupling between layers leads to a four-component charge-density wave forming nanostructures, as observed in a recent STM experiment[34]. This means that in-plane and out-of-plane orders are coupled and distorting one affects the other. While further experiments are needed to determine the precise origin of metastability, the present theoretical investigation reveals the importance of coupling between the in-plane and out-of-plane orders that gives rise to the complex free-energy landscape capable of hosting long-lived metastable states observed in switching experiments.

In conclusion, we have demonstrated non-volatile resistance switching between CDW states induced by electrical pulses across a broad temperature range that extends even above room temperature. Notably, switching allows us to reach states outside the hysteresis loop, revealing that the electrical pulse-excited state is a thermally inaccessible state. We find that the non-thermal switching mechanism is very energy efficient and requires only a low voltage for operation despite the large size of our devices. This compares favorably with other emerging memristor technologies[8,9] and highlights the potential of electronic switching between CDW states as a promising concept for ultra-efficient memristor applications. Our theoretical calculations propose an interlayer bistability mechanism in which the interlayer CDW distortion is fixed in tellurium bilayers and adopts two configurations in a single tellurium layer. This results in stable and metastable minima in the free-energy landscape, with their near-degeneracy lifted by the weak link provided by the mixed Eu–Te layer. A key future challenge is to understand the microscopic dynamics governing transitions between the two orientations and to identify the microscopic parameters that determine the switching time.

## Methods

**$EuTe_4$ crystal growth.** Single crystals of $EuTe_4$ were grown via a flux method. Europium pieces (99.99 %) and tellurium powder (99.999 %) were mixed in a ratio of 1:15 and put into a quartz ampule. The ampule was evacuated, sealed, and put into a cube furnace positioned nearly vertically. The furnace was heated over eight hours to reach 850 °C, then annealed for two days. After annealing, it was cooled over 100 hours to just above the melting point of the tellurium flux and held at this point for one week. The molten flux was then decanted, and the ampule was cooled. The resulting crystals are millimeter-sized and exhibit a golden, mirror-like surface. The crystal structure was confirmed by single-crystal X-ray diffraction.

**Device fabrication.** The circuits were fabricated on $Si/SiO_2$ substrates via standard optical lithography using the photoresists LOR3A and S1805. Metallic electrodes were made using an e-beam evaporator by first depositing 3 nm of Ti, followed by 100 nm of Au. Liftoff was performed in mr-rem 700, followed by rinsing in isopropanol. Device D1 was made by exfoliating the 80 nm thick and 4.4 μm wide $EuTe_4$ flake directly on the pre-fabricated circuit with gaps of 1.4 μm between the electrodes. This allowed us to minimize chemical processing for the device. Device D2 was fabricated using a standard procedure, which involved first exfoliating a 170 nm thick and 4.2 μm wide $EuTe_4$ flake on a $Si/SiO_2$ substrate, followed by the lithography procedure described above with gaps of 2.0 μm between the electrodes. The entire device fabrication process (except for laser patterning) was done in an $N_2$-filled glovebox to avoid device degradation. The flake dimensions were determined by an AFM. All device experiments were done in high vacuum. Both devices D1 and D2 show similar switching behavior.

**Device measurement and switching.** The resistance measurements of devices were done via a standard 4-probe technique with a Keithley 6221 current source and a Keithley 2182A voltmeter. Pulsed electrical experiments were done either with current pulses from the Keithley 6221 or by voltage pulses from the Siglent SDG1050 function generator. To minimize Joule heating during the *I-V* measurements, the current is applied in pulses while the voltage is measured during each pulse. Pulses are shorter than 1 ms and are applied with a period of 100 ms. Time-resolved measurements of voltage across the $EuTe_4$ device have been performed with an oscilloscope Siglent SDS5104X.

**Tight-binding model.** We describe the near-square Te layers in $EuTe_4$, which participate in the low-energy physics of interest[18], using a minimal tight-binding model[21,43]. We consider a square lattice with two orthogonal $p$-type orbitals per site ($p_x$ and $p_y$) at a filling of 2/3 and allow for nearest neighbour hopping. The in-plane electronic Hamiltonian is

$$H_{\parallel} = \sum_{\mathbf{k},\nu} \varepsilon_{\nu}(\mathbf{k}) c^{\dagger}_{\nu,\boldsymbol{k}} c_{\nu,\boldsymbol{k}} \tag{1}$$

where the orbital degrees of freedom are labelled by $\nu \in \{x, y\}$, and $c^{\dagger}_{\nu,\mathbf{k}}$ is the creation operator for the $p_{\nu}$ orbital at momentum $\mathbf{k}$. The electronic dispersion is given by

$$\varepsilon_{\nu}(\mathbf{k}) = -2t_{\sigma}\cos(k_{\nu}a) - 2t_{\pi}\cos(k_{\bar{\nu}}a)$$

where for $\nu = x$, we have $\bar{\nu} = y$ and vice-versa. The parameters used are $t_{\sigma} = -2$ eV for end-on hopping, $t_{\pi} = 0.2$ eV for side-on hopping, and $a = 3.2341$ Å for the lattice constant of the square Te layers (note that this is not the same as the lattice constant $\tilde{a}$ of the full $EuTe_4$ structure, which is given approximately as $\tilde{a} \approx \sqrt{2}a$).

**Interlayer hybridization.** We distinguish between an aligned bilayer configuration, where sites $(m, n)$ of the two layers are shifted by $(0,0,c)$, with layer spacing $c$, and a staggered bilayer configuration, where they are shifted by $(a/2, a/2, c)$. The interlayer coupling between neighbouring layers $l$ and $l + 1$ enters the electronic Hamiltonian as

$$H_{\perp} = \sum_{\mathbf{k},\nu} \left( t_{\perp}(\mathbf{k}) c^{\dagger}_{\nu,\mathbf{k},l} c_{\nu,\mathbf{k},l+1} + \text{h.c.} \right) \tag{2}$$

$$t_{\perp}(\mathbf{k}) = \begin{cases} -t_{\mathrm{al}} & \text{aligned configuration} \\ -t_{\mathrm{st}}\left(1 + e^{ik_x a}\right)\left(1 + e^{ik_y a}\right) e^{-\frac{i}{2}(k_x + k_y)a} & \text{staggered configuration} \end{cases} \tag{3}$$

For the energy surface calculation in Fig. 4a, we use a rough estimate[31] $t_{\mathrm{al}} = 0.1$ eV and $t_{\mathrm{st}} = 0.05$ eV, while for the Lindhard response shown in Fig. 4d, e, to better highlight the peak splitting, we use $t_{\mathrm{al}} = 0.3$ eV and $t_{\mathrm{st}} = 0.1$ eV.

**Lindhard response.** The optical susceptibility is computed on a discrete grid of $N_k$ momentum points for each of the $d$ dimensions using

$$\chi^R(\mathbf{q},\omega) = \frac{1}{{N_k}^2}\sum_{\mathbf{k}}\sum_{\nu,\nu'}\frac{f_\nu(\mathbf{k}+\mathbf{q}) - f_{\nu'}(\boldsymbol{k})}{\omega + i\eta + \varepsilon_\nu(\mathbf{k}+\mathbf{q}) - \varepsilon_{\nu'}(\mathbf{k})}\left\|\{U^\dagger(\mathbf{k}+\mathbf{q})\cdot U(\mathbf{k})\}_{\nu,\nu'}\right\|^2 \tag{4}$$

where $\nu$ and $n$ are orbital/band indices, is the corresponding electronic dispersion $f_\nu(\mathbf{k}) = f\big(\varepsilon_\nu(\mathbf{k})\big)$ is the Fermi-Dirac distribution function and $\cdot$ marks the matrix product. The electronic structure is obtained as an eigenvalue problem of the Hamiltonian matrix ${h^0}_{\alpha,\beta}(\mathbf{k}) = U_{\alpha,\nu}(\mathbf{k})\cdot\varepsilon_\nu(\mathbf{k})\cdot U^*_{\nu,\beta}(\mathbf{k})$, and the matrix $U(\mathbf{k})$ is a matrix of the column eigenvectors of said free Hamiltonian matrix $h^0(\mathbf{k})$. As an example, for a bilayer system (cf. Fig. 4d,e), we have

$$H^0(k) = \sum_{\mathbf{k},l} c^\dagger_{\boldsymbol{k},l}\, h^0(\boldsymbol{k})\, \boldsymbol{c}_{\mathbf{k},l} \tag{5}$$

$$= \sum_{\mathbf{k}}\begin{pmatrix} c^\dagger_{x,\mathbf{k},1} & c^\dagger_{y,\mathbf{k},1} & c^\dagger_{x,\mathbf{k},2} & c^\dagger_{y,\mathbf{k},2}\end{pmatrix}\begin{pmatrix} \varepsilon_x(\mathbf{k}) & 0 & t_\perp(\mathbf{k}) & 0 \\ 0 & \varepsilon_y(\mathbf{k}) & 0 & t_\perp(\mathbf{k}) \\ t_\perp(\mathbf{k}) & 0 & \varepsilon_x(\boldsymbol{k}) & 0 \\ 0 & t_\perp(\mathbf{k}) & 0 & \varepsilon_y(\boldsymbol{k})\end{pmatrix}\begin{pmatrix} c_{x,\mathbf{k},1} \\ c_{y,\mathbf{k},1} \\ c_{x,\mathbf{k},2} \\ c_{y,\mathbf{k},2}\end{pmatrix} \tag{6}$$

The broadening parameter is set to $\eta = 10^{-8}$ and we investigate the static response, i.e., $\omega = 0$.

**Lattice distortion and Peierls coupling.** The displacement of site $(m,n)$ is parametrised as a periodic lattice distortion (PLD) of the form

$$\boldsymbol{u}_{m,n} = \alpha\frac{2a}{\sqrt{3}}\begin{pmatrix}\sin\big(\mathbf{q}\cdot\boldsymbol{R}_{m,n} + \theta_x\big) \\ \sin\big(\mathbf{q}\cdot\boldsymbol{R}_{m,n} + \theta_y\big)\end{pmatrix} \tag{7}$$

with the CDW wave vector $\mathbf{q}$, equilibrium lattice positions $\mathbf{R}_{m,n} = a(m,n)$ [for staggered geometry $\mathbf{R}_{m,n} = a(m,n) + \frac{a}{2}(1,1)$], the amplitude $\alpha$ (given as a fraction of the lattice constant) and two phase parameters, $\theta_x$ and $\theta_y$. The electronic and lattice degrees of freedom are coupled via Peierls coupling, where the in-plane hopping is linearly suppressed with $\alpha$. This extends the electronic Hamiltonian with the following terms

$$H_{\text{el-lat}} = \sum_{\nu,m,n}\left[t_\sigma\,\lambda\nu_{m,n}\big(c^\dagger_{\nu,m,n}c_{\nu,m+1,n} + \text{h.c.}\big) + t_\pi\lambda\bar{\nu}_{m,n}\,\big(c^\dagger_{\nu,m,n}c_{\nu,m,n+1} + \text{h.c.}\big)\right] \tag{8}$$

where the magnitude of the displacement between neighboring sites, $\nu_{m,n}$ is either $\mathrm{x}_{m,n} = \|\mathbf{r}_{m+1,n} - \mathbf{r}_{m,n}\|$ or $\mathrm{y}_{m,n} = \|\mathbf{r}_{m,n+1} - \mathbf{r}_{m,n}\|$ for the 1st and 2nd term in $H_{\text{el-lat}}$, respectively with $\nu = \mathrm{x}$, and vice-versa with $\nu = y$ (and where $\mathbf{r}_{m,n} = \boldsymbol{R}_{m,n} + \boldsymbol{u}_{m,n}$) The coupling parameter is estimated by a qualitative fit to the experimental lattice distortion as $\lambda = 0.6$ Å$^{-1}$. We assume softening of a single phonon mode corresponding to a PLD commensurate with the square lattice, i.e., $\mathbf{q}_{\text{CDW}} = \frac{a^*}{3}(1,1)$. Moving to $k$-space and utilising the PLD parametrisation as a mean-field approximation, we obtain the effective electronic Hamiltonian for a single layer

$$H_{\mathrm{el}} = \sum_{\mathbf{k},\nu} c^{\dagger}_{\nu,\boldsymbol{k}} h_{\nu}(\mathbf{k}) c_{\nu,\mathbf{k}} \tag{9}$$

where

$$h_{\nu}(\boldsymbol{k}) = \begin{pmatrix} \varepsilon_{\nu}(\mathbf{k}) & \Delta^{*}_{\nu}(\mathbf{k},\mathbf{q}_{\mathrm{CDW}}) & \Delta_{\nu}\left(\mathbf{k}+2\mathbf{q}_{\mathrm{CDW}},\mathbf{q}_{\mathrm{CDW}}\right) \\ \Delta_{\nu}(\mathbf{k},\mathbf{q}_{\mathrm{CDW}}) & \varepsilon_{\nu}(\mathbf{k}+\mathbf{q}_{\mathrm{CDW}}) & \Delta^{*}_{\nu}(\mathbf{k}+\mathbf{q}_{\mathrm{CDW}},\mathbf{q}_{\mathrm{CDW}}) \\ \Delta^{*}_{\nu}(\mathbf{k}+2\mathbf{q}_{\mathrm{CDW}},\mathbf{q}_{\mathrm{CDW}}) & \Delta_{\nu}(\mathbf{k}+\mathbf{q}_{\mathrm{CDW}},\mathbf{q}_{\mathrm{CDW}}) & \varepsilon_{\nu}(\mathbf{k}+2\mathbf{q}_{\mathrm{CDW}}) \end{pmatrix} \tag{10}$$

$$c_{\nu,\mathbf{k}} = \begin{pmatrix} c_{\nu,\mathbf{k}} \\ c_{\nu,\mathbf{k}+\mathbf{q}_{\mathrm{CDW}}} \\ c_{\nu,\mathbf{k}+2\mathbf{q}_{\mathrm{CDW}}} \end{pmatrix} \tag{11}$$

and the CDW order parameter is

$$\Delta_{\nu}(\mathbf{k},\mathbf{q}) = g_{\sigma}(k_{\nu},q_{\nu})u_{\mathbf{q},\nu} + g_{\pi}(k_{\bar{\nu}},q_{\bar{\nu}})u_{\mathbf{q},\bar{\nu}} \tag{12}$$

with the momentum-dependent electron-phonon coupling

$$g_{\sigma}\left(k_{\nu},q_{CDW,\nu}\right) = \frac{2i\lambda t_{\sigma}}{N}\left[\sin\left(\left(k_{\nu}+q_{\mathrm{CDW},\nu}\right)a\right) - \sin(k_{\nu}a)\right] = -i\frac{2\sqrt{3}\lambda}{N}t_{\sigma}\cos\left(k_{\nu}a+\frac{\pi}{3}\right) \tag{13}$$

and canonical displacement variables $u_{\boldsymbol{q},x}$ and $u_{\boldsymbol{q},y}$ being the sole non-zero Fourier coefficients of the real-space displacement of the lattice sites, i.e., $\boldsymbol{u}_{m,n} = \sum_{\mathbf{q}} u_{\mathbf{q}} e^{i\mathbf{q}\cdot\mathbf{R}}$.

The lattice energy per site is computed using the harmonic approximation as $H_{\mathrm{lat}} = 2Ka^2\alpha^2$, with the lattice constant estimated[28] as $K = 1.44$ eV/Å$^2$. The dimensionless parameter quantifying the strength of the electron-lattice response is thus $t_{\sigma}\lambda/Ka \approx 0.26$.

**Numerical simulations.** The numerical simulations were implemented in Python with the aid of the NumPy and SciPy modules. A $k$-point grid of $200 \times 200$ points was determined to be sufficiently accurate for the energy surface calculations, while a $300 \times 300$ grid was used for computing the Lindhard response. The optimization w. r. t., $\left(\alpha,\theta_x,\theta_y\right)$ was performed by using the modified Powell method to obtain the roots of the equilibrium condition equation, which in turn was derived using the Hellmann-Feynman theorem. The accuracy threshold for the PLD parameters was set to $\epsilon = 10^{-5}$. It is important to emphasize that the degeneracy for the monolayer, Fig. 4a, arises only when each configuration is relaxed with respect to the phase of the charge-density wave in both the *x* and *y* directions at a given distortion $\alpha$.

## Data Availability

The data that support the findings of this Article are available via Zenodo at https://zenodo.org/records/19683604

## Code Availability

Code used for the calculations supporting our findings is available via Zenodo at https://zenodo.org/records/17908807

**Acknowledgments**

We thank T. Mertelj, I. Vaskivskyi, V. Kabanov, and A. Mraz for fruitful discussions. We thank D. Vengust and E. Goreshnik for their assistance with crystal characterization.

**Funding**

D. M. wishes to acknowledge funding from ERC AdG 'HIMMS' and the Slovenian Research and Innovation Agency (ARIS) for programs N1-0290 and N1-0295. R. V., Y. V., and D. M. thank ARIS for funding the research program P1-0040. R. V. acknowledges funding from the Swiss National Science Foundation (SNSF) and ARIS as a part of the WEAVE framework Grant Number 213148. D. Golež acknowledges the support of programs No. P1-0044, No. J1-2455 and MN-0016-106 of the ARIS.

**Author Contributions**

D. M. and D. Golež supervised the project. R. V. conceived the experiments. P. S. grew $EuTe_4$ crystals. R. V., M. R., and J. L. fabricated the devices. R. V., M. R., and J. L. conducted experiments with support from Y. V., F. Š. and D. Grabnar. J. G. and D. Golež performed theoretical calculations. R. V. and D. Golež wrote the paper with input from all authors.

**Competing interests**

The Authors declare no Competing Financial or Non-Financial Interests.